\documentclass{article}
\usepackage{graphicx}
\usepackage{amsmath}
\usepackage{amsfonts}
\usepackage{amssymb}
\newtheorem{theorem}{Theorem}

\begin{document}

\title{Facts and Fictions about Anti de Sitter Spacetimes with Local Quantum Matter\\{\small Dedicated to the memory of Harry Lehmann}}
\author{Bert Schroer\\Institut f\"{u}r Theoretische Physik\\FU-Berlin, Arnimallee 14, 14195 Berlin, Germany\\presently: CBPF, Rua Dr. Xavier Sigaud, 22290-180 Rio de Janeiro, Brazil \\schroer@cbpf.br}
\date{November 1999}
\maketitle
\begin{abstract}
It is natural to analyse the AdS$_{d+1}$-CQFT$_{d}$ correspondence in the
context of the conformal- compactification and covering formalism. In this way
one obtains additional inside about Rehren's rigorous algebraic holography in
connection with the degree of freedom issue which in turn allows to
illustrates the subtle but important differences beween the original string
theory-based Maldacena conjecture and Rehren's theorem in the setting of an
intrinsic field-coordinatization-free formulation of algebraic QFT. I also
discuss another more generic type of holography related to light fronts which
seems to be closer to 't Hooft's original ideas on holography. This in turn is
naturally connected with the generic concept of ``Localization Entropy'', a
quantum pre-form of Bekenstein's classical black-hole surface entropy.
\end{abstract}

\section{Historical background}

There has been hardly any problem in particle physics which has attracted as
much attention as the problem if and in what way quantum matter in the
\textbf{A}nti \textbf{d}e\textbf{S}itter spacetime and the one dimension lower
conformal field theories are related and whether this could possibly contain
clues about the meaning of quantum gravity.

In more specific quantum physical terms the question is about a conjectured
\cite{Ma}\cite{GKP}\cite{Wi} (and meanwhile in large parts generically and
rigorously understood \cite{Re}) correspondence between two quantum field
theories in different spacetime dimensions; the lower-dimensional conformal
one being the ``holographic image'' or projection of the AdS theory.
Conjectures, different from mathematical proofs; allow of course almost always
a certain margin in their precise mathematical formulation and in their
physical interpretation. The field theoretic content of this conjecture has
often been interpreted as a correspondence between two Lagrangian field
theories (e.g. between a conformally invariant 4-dimensional SYM and a higher
dimensional spin=2 gravitational-like theory). The exact theorem says that
such a correspondence cannot exist; one side has to be non-Lagrangian. There
is no exception to this proposition; not even the assumption of supersymmetry
helps here. One of our goals is to spell this out in detail and to illustrate
this interesting point with a simple model.

The community of string physicists has placed this correspondence problem in
the center of their interest. Remembering the great conceptual and
calculational achievements as e.g. the derivation of scattering theory and
dispersion relations from field theory with which the name of Harry Lehmann
(to whose memory this article is dedicated) is inexorably linked, I will limit
myself to analyze the particle physics content of the so-called Anti deSitter
conformal QFT-correspondence from the conservative point of view of a quantum
field theorist who, although having no active ambitions outside QFT, still
nourishes a certain curiosity about present activities in particle physics as
e.g. string theory or noncommutative geometry. In the times of Harry Lehmann
the acceptance of a theoretical proposal in particle physics was primarily
coupled to its experimental verifiability and/or its conceptual standing
within physics.

The AdS model of a curved spacetime has a long history \cite{Frons}\cite{AIS}
as a theoretical laboratory of what can happen with particle physics in a
universe which is the extreme opposite of globally hyperbolic in that it
possesses a self-closing time, whereas the proper de Sitter spacetime was once
considered among the more realistic models of the universe. The recent surge
of interest about AdS came from string theory and is different in motivation
and more related to the hope (or dream) to attribute a meaning to ``Quantum
Gravity'' from a string theory viewpoint.

Fortunately for a curious outsider (otherwise I would have to quit right
here), this motivation has no bearing on the conceptual and mathematical
problems posed by the would be AdS-conformal QFT correspondence; the latter
turned out to be one of those properties discovered in the setting of string
theory which allow an interesting and rigorous formulation in QFT which
confirms some, but not all the conjectured properties.

The rigorous treatment however requires a reformulation of (conformal) QFT
within a more algebraic setting. The standard formalism based on pointlike
``field coordinatizations'' which underlies the Lagrangian (and the Wightman)
fomulations does not provide a natural setting for the study of isomorphisms
between models in different spacetime dimensions, even though the underlying
physical principles are the same. One would have to introduce many additional
concepts and auxiliary tricks into the standard framework to the extend that
the formulation appears contrived containing too many ad hoc prescriptions.
The important aspects in this isomorphism are related to space and time-like
(Einstein,Huygens) causality, localization of corresponding objects and
problems of degree of freedom counting. All these issues are belonging to
real-time physics and in most cases their meaning in terms of Euclidean
continuation (statistical mechanics) remains obscure; but this of course does
not make them less physical.

This note is organized as follows. In the next section I elaborate the
kinematical aspects of the AdS$_{d+1}$-CQFT$_{d}$ situation as a collateral
result of the old (1974/75) compactification formalism for the
``conformalization'' of the d-dimensional Minkowski spacetime. For this reason
the seemingly more demanding problem of studying QFT directly in AdS within a
curved spacetime formalism\footnote{This was also done in the 70ies by
Fronsdal. There was a good reason why he missed the isomorphism to CQFT
despite his musterful handling of (noncompact) group theory: it was the degree
of freedom (multiplicity) problem which will be addressed later.} can be
bypassed. The natural question whose answer would have led directly from
$CQFT_{4}$ to $AdS_{5}\,$\ in the particle physics setting (without string
theory as a midwife) is: does there exist a quantum field theory which has the
same $SO(4,2)$ symmetry and just reprocesses the $CQFT_{4}$ matter content in
such a way that the ``conformal Hamiltonian'' (the timelike rotational
generator through compactified $\bar{M})$ becomes the true hamiltonian? This
theory indeed exists, it is an AdS theory with a specific local matter content
computable from the CQFT matter content. The answer is unique, but as a result
of the different dimensionality one cannot describe this one-to-one relation
between spacetime indexed matter contents in terms of pointlike fields. This
will be treated in section 3, where we will also compare the content of
Rehren's isomorphism \cite{Re}\cite{BFS} with the Maldacena, Witten at al.
\cite{Ma}\cite{GKP}\cite{Wi} conjectures and notice some subtle but
potentially serious differences in case one interpretes the conjecture (as it
was done in most of the subsequent literature) as a relation between two
Lagrangian theories. Who is aware of the fact that subtle differences often
have been the enigmatic motor of progress, will not dismiss such observations.

The last section presents some general results of AQFT on
degrees-of-freedom-counting and holography. Closely connected is the idea of
``chiral scanning'' i.e. the encoding of the full content of a higher
dimensional (massive) field QFT into a finite number of copies of one chiral
theory in a carefully selected relative position within a common Hilbert
space. In this case the prize one has to pay for this more generic holography
(light-front holography) is that some of the geometrically acting spacetime
symmetry transformations become ``fuzzy'' in the holographic projection and
some of the geometrically acting symmetries on the holographic image are not
represented by diffeomorphisms if pulled back into the original QFT.

\section{ Conformal Compactification and AdS}

The simplest type of conformal QFT is obtained by realizing that zero mass
Wigner representation of the Poincar\'{e} group with positive energy (and
discrete helicity) and allow for a natural extension to the conformal symmetry
group $SO(4,2)/Z_{2}$ \textit{without any enlargement of the Hilbert space}.
Besides scale transformations, this larger symmetry also incorporates the
fractional transformations (proper conformal transformations)
\begin{equation}
x^{\prime}=\frac{x-bx^{2}}{1-2bx+b^{2}x^{2}} \label{sc}%
\end{equation}
It is often convenient to view this formula as the action of the translation
group $T(b)$ conjugated with a (hyperbolic) inversion $I$
\begin{align}
I  &  :x\rightarrow\frac{-x}{x^{2}}\\
x^{\prime}  &  =IT(b)Ix
\end{align}
$I$ does not belong to the above conformal group, although it is unitarily
represented (and hence a Wigner symmetry) in these special Wigner
representations. For fixed $x$ and small $b$ the formula (\ref{sc}) is well
defined, but globally it mixes finite spacetime points with infinity and hence
requires a more precise definition in particular in view of the positivity
energy-momentum spectral properties in its action on quantum fields. Hence as
preparatory step for the adequate formulation of quantum field theory
concepts, one has to achieve a geometric compactification. This starts most
conveniently from a linear representation of the conformal group $SO(d,2)$ in
d+2-dimensional auxiliary space $\mathbb{R}^{(d,2)}$ (i.e. without field
theoretic significance) with two negative (time-like) signatures
\begin{equation}
G=\left(
\begin{array}
[c]{ccc}%
g_{\mu\nu} &  & \\
& -1 & \\
&  & +1
\end{array}
\right)
\end{equation}
and restricts this representation to the $(d+1)$-dimensional forward light
cone
\begin{equation}
LC^{(d,2)}=\{\xi=(\mathbf{\xi},\xi_{4},\xi_{5});\mathbf{\xi}^{2}+\xi_{d}%
^{2}-\xi_{d+1}^{2}=0\}
\end{equation}
where $\mathbf{\xi}^{2}=\xi_{0}^{2}-\vec{\xi}^{2}$ denotes the d-dimensional
Minkowski length square. The compactified Minkowski space $\bar{M}_{d}$ is
obtained by adopting a projective point of view (stereographic projection)
\begin{equation}
\bar{M}_{d}=\left\{  x=\frac{\mathbf{\xi}}{\xi_{d}+\xi_{d+1}};\xi\in
LC^{(d,2)}\right\}
\end{equation}
It is then easy to verify that the linear transformation, which keep the last
two components invariant consist of the Lorentz group and those
transformations which only transform the last two coordinates, yield the
scaling formula
\begin{equation}
\xi_{d}\pm\xi_{d+1}\rightarrow e^{\pm s}(\xi_{d}\pm\xi_{d+1})
\end{equation}
leading to $x\rightarrow\lambda x,\lambda=e^{s}\,.$ The remaining
transformations, namely the translations and the fractional proper conformal
transformations, are obtained by composing rotations in the $\mathbf{\xi}_{i}%
$-$\xi_{d}$ and boosts in the $\mathbf{\xi}_{i}$-$\xi_{d+1}$ planes.

A convenient description of Minkowski spacetime $M$ in terms of this $d+2$
dimensional auxiliary formalism is obtained in terms of a ``conformal time''
$\tau$%
\begin{align}
M_{d}  &  =(sin\tau,\mathbf{e,}cos\tau),\,\,e\in S^{d-1}\\
t  &  =\frac{sin\tau}{e^{d}+cos\tau},\,\,\vec{x}=\frac{\vec{e}}{e^{d}+cos\tau
}\label{M}\\
e^{d}+cos\tau &  >0,\,\,-\pi<\tau<+\pi\nonumber
\end{align}
so that the Minkowski spacetime is a piece of the d-dimensional wall of a
cylinder in d+1 dimensional spacetime which becomes tiled with the closure of
infinitely many Minkowski worlds. If one cuts the wall on the backside
appropriately, this carved out piece representing d-dimensional compactified
Minkowski spacetime has the form of a d-dimensional double cone symmetrically
around $\tau=0,\mathbf{e=(0},e^{d}=1\mathbf{)}$ without its
boundary\footnote{The graphical representations are apart from the
compactification (which involves identifications between past and future
points at time/light-infinity) the famous Penrose pictures of $M.$}%
$\mathbf{.}$ The above directional compactification leads to an identification
of boundary points at ``infinity'' and give e.g. for d=1+1 the compactified
manifold the topology of a torus. The points which have been added at the
infinity to $M$ namely $\bar{M}\backslash M$ are best described in terms of
the d-1 dimensional submanifold of points which are lightlike with respect to
the past infinity apex at $m_{-\infty}=(0,0,0,0,1,\tau=-\pi)$. The cylinder
walls form the universal covering $\widetilde{M_{d}}=S^{d-1}\times\mathbb{R}$
which is ``tiled'' in both $\tau$-directions by infinitely many Minkowski
spacetimes (``heavens and hells'') \cite{LM}. If the only interest would be
the description of the compactification $\bar{M},$ then one may as well stay
with the original x-coordinates and write the d+2 $\xi$-coordinates follow
Dirac and Weyl as%

\begin{align}
&  \xi^{\mu}=x^{\mu},\,\,\mu=0,1,2,3\\
&  \xi^{4}=\frac{1}{2}(1+x^{2})\nonumber\\
&  \xi^{5}=\frac{1}{2}(1-x^{2})\nonumber\\
&  i.e.\,\,\,\left(  \xi-\xi^{\prime}\right)  ^{2}=\left(  x-x^{\prime
}\right)  ^{2}\nonumber
\end{align}
Since $\xi$ is only defined up to a scale factor, we conclude that lightlike
differences retain an objective meaning in $\bar{M}$ even though the space-
and time-like separation does loose its meaning. An example of a physical
theory on $\bar{M}$ are free photons. The impossibility of a distinction
between space- and time- like finds its mathematical formulation in the
Huygens principle which says that the lightlike separation is the only one
where the physical fields do not commute and hence where an interaction can
happen. In the terminology of local quantum physics this means that the
commutant of an observable algebra localized in a double cone consists
apparently of a (Einstein causal) connected spacelike- as well as two
disconnected (Huygens causality) timelike- pieces. But taking the
compactification into consideration one realizes that all three parts are
connected and the space/time-like distinction is meaningless on $\bar{M}.$ In
terms of Wightman correlation functions this is equivalent to the rationality
of the analytically continued Wightman functions of observable fields which
includes an analytic extension into timelike Jost points \cite{To1}%
\cite{anomalous}.

Therefore in order to make contact with particle physics aspects, the use of
either the covering $\widetilde{M}$ or of more general fields (see next
section) on $\bar{M}$ is very important since only in this way one can
implement the pivotal property of causality together with the associated
localization concepts. As first observed by I. Segal \cite{Se} and later
elaborated and brought into the by now standard form in field theory by
L\"{u}scher and Mack \cite{LM}, a global form of causality can be based on the
sign of the invariant%

\begin{align}
&  \left(  \xi(\mathbf{e},\tau)-\xi(\mathbf{e}^{\prime},\tau^{\prime})\right)
^{2}\gtrless0,\,\,hence\\
&  \left|  \tau-\tau^{\prime}\right|  \gtrless2\left|  Arcsin\left(
\frac{\mathbf{e}-\mathbf{e}^{\prime}}{4}\right)  ^{\frac{1}{2}}\right|
=\left|  Arccos\left(  \mathbf{e\cdot e}^{\prime}\right)  \right|  \nonumber
\end{align}
where the $<$ inequality characterizes global spacelike distances and $>$
corresponds to positive and negative global timelike separations. Whereas the
globally spacelike region of a point is compact, the timelike region is not.
The concept of global causality solves the so called Einstein causality
paradox of CQFT \cite{HSS}. In the next section we will meet a global
decomposition method which also avoids this paradox without the necessity \ of
using covering space.\ 

The central theme, namely the connection with QFT on AdS enters this section
naturally if one asks the question whether one can instead of the surface of
the forward light cone alternatively use a mass hyperboloid $H_{d+1}$ inside
the forward light cone of the same ambient d+2 dimensional space%
\begin{align}
H_{d+1}  &  =\left\{  \eta;\eta^{2}=1\right\} \\
\eta^{0}  &  =\sqrt{1+r^{2}}sin\tau\nonumber\\
\eta^{i}  &  =r\mathbf{e}^{i},\,\,i=1,...d\nonumber\\
\eta^{d+1}  &  =\sqrt{1+r^{2}}cos\tau\nonumber
\end{align}
This space which because of its formal relation to the analogous deSitter
spacetime (which is defined by the spacelike hyperboloid) is called ``Anti
deSitter'' spacetime is noncompact. It is obvious from its construction that
its asymptotic part is the same as $\bar{M}_{d}.$ It was conjectured by
Maldacena and others \cite{Ma}\cite{GKP}\cite{Wi} that there is also a
correspondence between quantum field theories. This conjecture implies the
tacit assumption (not explicitly stated in these papers) that an $AdS_{d+1}$
QFT which coalesces asymptotically\footnote{Using the previous cylindric
representation of the conformal covering, the covering of AdS corresponds to
the full cylinder of which its mantel is the conformal covering.} with an
$CQFT_{d}$ theory has a unique extension into the AdS bulk. Since there can be
no mapping between pointlike fields on spacetimes of different dimensions the
question of the origin of this unique extension is non-trivial. The conjecture
came from some speculations concerning possible relations of string theory
with some supersymmetric gauge theories (SYM) i.e. from ideas far removed from
the present particle physics setting which will not be explained here.

In the 70s, at the time of the conformal compactifications, free fields on
$AdS_{4}$ were studied from a particle physics viewpoint by Fronsdal
\cite{Frons}. The correspondence to $CQFT_{3}$ was overlooked; probably
because of the fact that despite the obvious group theoretical connection
through the common $SO(3,2),$ the multiplicites of the discrete AdS free
Hamiltonian turned out too big for matching those of the rotational conformal
Hamiltonian; a fact which will find its explanation in the next section.

Although the two spacetime cannot be mapped into each other, their shared
spacetime symmetry group $SO(4,2)$ suggests that there is at least a
correspondence between certain subsets which may be obtained from projecting
down wedge regions from the ambient space onto the two spacetime manifolds.
Wedges have a natural relation to $SO(4,2)$; they may all be generated from
standard wedge in the ambient auxiliary space $W_{st}=\left\{  \xi^{1}>\left|
\xi^{0}\right|  \right\}  .$ The fixed point group of this transitive action
on wedges consists of a boost and transversal translations and
rotations\footnote{If one adds the two longitudinal lightlike translations
which in one direction cause a compression into the wedge, one obtains a
8-dimensional Galilei group \cite{SW}.
\par
{}
\par
{}}. The projected wedges $pW$ on AdS are by definition again wedges in
AdS/CQFT and the $SO(d,2)$ symmetry group has the same transitive action i.e
the system of wedges is described by $SO(d,2)$ modulo the fixed point
subgroup. This geometric situation clearly suggests that on should consider
algebras associated with these wedges instead of looking for a relation
between pointlike fields. On the conformal side this includes all double cone
algebras of arbitrary small size since the noncompact wedge regions are
conformally equivalent to compact double cones regions. The logic of algebraic
QFT requires to continue this algebraic correspondence to all intersections
obtained from wedges. In this way one expects to arrive at an isomorphism
which carries the full content of both theories and which includes the
asymptotic relation (on the conformal surface of the aforementioned cylinder)
in terms of field coordinatizations used by Maldacena et al. In order to
obtain a rigorous proof, one must check some consistency conditions in the
conversion of maps between spacetime regions and algebras indexed by those
regions. This was achieved by Rehren \cite{Re} and will be briefly comment on
his theorem (including its relation to the original conjecture) in the next section.

According to our previous remarks, interacting conformal local fields live on
the covering space $\widetilde{M}.$ Fortunately the geometric isomorphism
between wedge regions can be lifted to an $\widetilde{AdS}_{d+1}-\widetilde
{M}$ correspondence. The conformal decomposition theory of the next section
avoids the use of the rather complicated coverings by using an operator analog
of fibre bundles on $\bar{M}$

\section{The conformal Hamiltonian as the true Hamiltonian}

There is another less geometric, but more particle physics type of argument,
which leads to the AdS-CQFT correspondence.

For this one should recall that in $SO(d,2)$ there are besides the usual
translations with infinity as a fixed point also ``conformal translations''
which act on the compactified $\bar{M}$ without fixed points as some kind of
``timelike rotations''. They are the analogs of the light-like chiral rotation
$R^{(\pm)}$ ($L_{0}^{(\pm)}$ in standard Virasoro algebra notation) and their
connection with the light ray translation $P^{(\pm)}$ with which they share
the positivity of their spectrum is
\begin{align}
R^{(\pm)}  &  =P^{(\pm)}+K^{(\pm)}\\
K^{(\pm)}  &  =I^{(\pm)}P^{(\pm)}I^{(\pm)}\nonumber
\end{align}
where $I_{\pm}$ is the representer of the chiral conformal reflection
$x\rightarrow-\frac{1}{x}$ (in linear lightray coordinates $x$) and $K$ is the
generator of the fractional special conformal transformation (\ref{sc}). For
free zero mass fields the discrete $R$-spectrum can be understood in terms of
that of a Hamiltonian for a massless model in a spatial box. This is however
not possible for the $R$-spectrum of chiral theories with anomalous scale
dimension (the R-spectrum is known to be identical to the that of scale
dimensions). In that case the only theory for which the spectrum is that of
its Hamiltonian is the QFT on $AdS_{2}.$ So if one wants to read the $SL(2,Z)$
modular characters of chiral conformal field theory in the spirit of a
Hamiltonian Gibbs formula one should use the AdS side. An analogous statement
holds in higher dimensions where the $\bar{M}$ rotation is described in terms
of a Lorentz vector $R_{\mu}$
\begin{equation}
R_{\mu}=P_{\mu}+IP_{\mu}I\nonumber
\end{equation}
where the inversion $I$ was defined at the beginning of the previous section.
It leads to a family of operators with discrete spectrum of $e\cdot R$ which
are dependent on a timelike vector $e_{\mu}.$ Again the operator $R_{0}$ is
the true Hamiltonian of only one theory with the same symmetry group and the
same system of algebras (but with a different spacetime indexing): the
associated d+1 dimensional $AdS$ theory.

Now it is time to quote (adapted to our purpose) Rehren's theorem and comment
on it.

\begin{theorem}
The geometric bijection between projected wedges pW on AdS$_{d+1}$ and the
conformal double cones in \={M}$_{d}$ which constitute the asymptotic infinity
of pW (as described in the previous section) extends to an isomorphism of the
corresponding algebras. Both theories share the same Hilbert space and the
same family of operator algebras but their spacetime organization and with it
their physical interpretation changes.
\end{theorem}

For the proof we refer to Rehren \cite{Re}.

Some comments are in order. There is no additional restrictive assumptions
(supersymmetry, vanishing $\beta$-functions) on either side. If the algebras
of the AdS theory are generated by pointlike fields then the associated
conformal algebra cannot be generated by a field which has an energy-momentum
tensor or obeys a causal equation of motion. This is one of Rehren's
conclusions and it is very instructive to illustrate this with an example.

Consider a free scalar AdS field \cite{Ber}. A simple calculation which will
not be repeated here reveals that it corresponds to a conformal
\textit{generalized free field} with homogeneous Kallen-Lehmann spectral
function. Generalized free fields always have been physically suspect and if
there spectral functions increases in the manner as the homogeneous degree
demands in this case, one can even prove that primitive causality \cite{HS} is
violated since the algebra on a piece of time-slice (represented as a chain of
small double cones which approximate the compact slice ffrom the inside) is
not equal to its causal completion (causal shadow) algebra. As one moves up in
time inside the causal shadow from the time-slice more and more degrees of
freedom coming from the inner parts of the bulk enter the causal shadow which
were not in the time-slice. Rehren's graphical representation \cite{Reh2} of
the CQFT world on the wall of a full AdS cylinder makes this undesired
sidewise propagation geometrically visible. This free field situation is
generic in the sense that pointlike AdS fields always carry \textit{too many
degrees of freedom }which leads to a violation of causal propagation in the
aforementioned sense\footnote{Contrary to a widespread belief, the number of
degrees of freedom of causally propagating AdS theories is always larger than
that of causally propagating conformal theories so that the isomorphism cannot
be one among causally propagating theories. If the AdS theory is pointlike and
causally propagating, the associated conformal theory has no causal
propagation.}. Such theories have to be abandoned for general physical reasons
(not just because they do not fit into a Lagrangian picture which
automatically implies causal propagation). Therefore the nice idea
\cite{Ruehl} to circumvent the scarcity in constructing Lagrangian conformal
models (the $\beta$-function restrictions) by starting instead with AdS
Lagrangians does not work, since the resulting conformal theories all share
the above defect.

In passing we mention that the brane idea shares the same causality conflict
with pointlike field. Whereas from a mathematical viewpoint a manifold of
interest may in certain cases be considered a brane of a larger dimensional
space, the assignment of a physical reality to the ambient spacetime generates
causality problems of the above kind for restrictions to the brane in case of
pointlike field theories in the larger ambient spacetime. Only if the ambient
degrees of freedom are carefully tuned to the brane can such causality
violations be avoided. Note that it is always the causal shadow property which
may get lost in such constructions and not the Einstein causality. This is not
visible if one restricts ones attention to (semi)classical solutions
concentrated on a brane and or to euclidean formulations. Whereas the
principles of AQFT confirm in a very precise way that there exists an
isomorphism it is very interesting that there is a clash with certain concepts
which have been used in string theory for the last two decades. This clash
extents beyond the above remarks on the AdS-CQFT and the brane concept and
casts doubt on the consisteny of such quasiclassical pictures as the
Kaluza-Klein dimensional reduction.

As a matter of fact not even the quasiclassical Klein-Kaluza reduction idea
has been shown to be consistent with causal QFT. For this one would have to
demonstrate that the idea works on the ready made QFT and not just on the
objects involved in the formal quantization approach which is used in the
tentative construction of a QFT. As far as the strict conceptual requirement
of causality and Haag duality in AQFT are concerned, the K-K mechanism, to the
extend that it is not just a mathematical trick (but an asymptotic property of
a genuine inclusion of two local quantum physics worlds) has at best remained
an enigmatic speculative idea (and at worst a tautology caused by not doing
what one actually is claiming to do).

The above degree of freedom discussion creates the suspicion that ``good''
causal conformal theories may have too few degrees of freedom in order to
yield AdS pointlike fields as the other side of the coin of the above
observation that pointlike AdS fields create causally bad conformal theories
This is indeed the case and can be seen by starting from the Wigner zero mass
representation space of the Poincar\'{e} group%
\begin{equation}
H_{Wig}=\left\{  \psi(\vec{p})|\int\left|  \psi(\vec{p})\right|  ^{2}%
\frac{d^{d-1}p}{2\left|  \vec{p}\right|  }<\infty\right\}
\end{equation}
which \textit{without extension} provides an irreducible representation space
of $SO(d,2).$ The subspace of modular wedge-localized Wigner wave functions
consists of boundary values of wave functions $\psi$ which are analytic in the
rapidity strip $0<Im\theta<\pi$ where for the standard wedge%
\begin{equation}
p_{x}=\sqrt{p_{y}^{2}+p_{z}^{2}}sinh\theta,\,p_{0}=\sqrt{p_{y}^{2}+p_{z}^{2}%
}cosh\theta
\end{equation}
This wave function space is in common regardless whether we are talking about
the standard wedge on $M_{d}$ or the corresponding $AdS_{d+1}$ wedge. Whereas
this space in the $M_{d}$ interpretation is easily rewritten in terms of
covariant x-space wave functions with the expected support properties, an
analog $\eta$-spacetime covariantization for AdS does not work. The best one
can do is to introduce (Olsen-Nielsen) string-like wave functions in $\eta
$-spacetime which do not depend on $w$ and which behave under $SO(d,2)$
transformations as objects which depend on an additional direction (the string
direction). So instead of pointlike fields one obtains \textit{strand-like
objects} (with weaker covariance properties) which emanate from the points on
the asymptotic boundary and extend through the bulk and which are linear in
the same momentum space creation/annihilation operators as those which appear
in the free conformal fields. This is the way in which the \textit{AdS
formulation maintains the conformal degrees of freedom and the primitive
causality}.

At this point an ardent string theorist might say: didn't I tell you that
starting from a conformal field theory you should expect to encounter AdS
strings! However the strand-like objects of the Rehren theorem \cite{Re} are
perfectly within causal localizable AQFT, and hence they are not objects of
string theory proper. The main characteristics of the strings of string theory
namely the enhancement of the degrees of freedom due to the internal
excitation structure is missing in the case of our strands. In order to avoid
misunderstandings we emphasize again that the degree of freedom issue is not
related to Einstein causality which remains valid irrespective of whether the
local algebras are generated by pointlike fields or not, but rather to the
causal propagation property which requires Einstein causality as a
prerequisite, but is not guarantied by the lattter.

The Maldacena et. al conjecture \cite{Ma}\cite{GKP}\cite{Wi} is that some high
dimensional true (i.e. not the above kinematical strands) string theories in
some effective and not precisely specified sense is equivalent with
conformally invariant supersymmetric Yang-Mills (SYM) theories. To the extend
in which the argument supporting this conjecture uses a correspondence between
pointlike (Lagrangian) $AdS_{5}$ QFT and a conformal SYM it is contradicted by
the above theorem.

Antinomies and contradictions about important topics in earlier times were
often the source of progress and removal of prejudices and one would hope that
they continue to receive their due attention. The issue is somewhat delicate
as a result that the euclidean functional integral formalism in which the
original conjectures were presented is at most an heuristic starting point
since Feynman-Kac representations in strictly renormalizable QFTs \textit{are
simply not valid} for the physical (renormalized) results. For this reason the
chosen method poses an obstacle against converting the conjecture into a
proof. This conceptual flaw of the functional action approach is one of the
raisons d'etre of algebraic QFT which succeeds to balance the starting
calculational definitions with the properties of the constructed models. One
could of course try to argue exclusively in terms of differential geometric
concepts by abstacting from the action formulation \textit{purely geometric}
definitions of what constitutes SYMs and the associated string theories. But
in doing this one will lose the relation to local quantum physics and the
obtaines theorem may be void of an particle physics content.

If one admits that, as argued above, the Lagrangian perturbation method
applied to the AdS side of the correspondence cannot be used for the
construction of additional conformal QFTs, the question arises whether there
are other construction methods. A closer investigation reveals that the
spectrum of anomalous dimensions of interacting conformal field theories is
determined in terms of a timelike braid group structure. A convenient way of
presenting this structure is to work with nonlocal component fields \cite{SS}
which result from the decomposition of the charge carrying globally local
fields $F$ on $\widetilde{M}$ under the reduction of the center of the
conformal covering group $(\widetilde{S(D,2)})$
\begin{align}
F(x)  &  =\sum_{\alpha,\beta}F_{\alpha,\beta}(x),\,\,F_{\alpha,\beta}(x)\equiv
P_{\alpha}F(x)P_{\beta}\\
Z  &  =\sum_{\alpha}e^{2\pi i\theta_{\alpha}}P_{\alpha}\nonumber
\end{align}
in terms of projectors $P_{a}$ which appear in the spectral decomposition of
the generator $Z$ of the $center(\widetilde{S(D,2)})=\left\{  Z^{n}%
,n\in\mathbb{Z}\right\}  $. In a way the existence of this decomposition
facilitates the use of the standard parametrization of Minkowski space
augmented by the quasiperiodic central transformation
\begin{equation}
ZF_{\alpha,\beta}(x)Z^{\ast}=e^{2\pi i(\theta_{\alpha}-\theta_{\beta}%
)}F_{\alpha,\beta}(x)
\end{equation}
and hence one may to a large part avoid the use of the complicated covering
parametrization and its $\widetilde{SO(D,2)}$ transformations which the
unprojected fields $F$ would require. For the latter fields on $\widetilde{M}$
the notation would be insufficient; one also has to give an equivalence class
of path (the number $n\gtrless$0\ of the heaven/hell one is in) with respect
to our copy of $M$ embedded in $\widetilde{M}.$ The projected fields on the
other hand are analogous to sections in a trivialized vectorbundle. With the
help of conformal 3-point functions one shows that the $\theta$-phases are
related to the anomalous dimensions. The component fields $F_{\alpha,\beta
}(x)$ are the suitable objects for the formulation of the timelike braid group
commutation relations which take the form of an exchange algebra%

\begin{equation}
F_{\alpha,\beta}(x)G_{\beta,\gamma}(y)=\sum_{\beta^{\prime}}R_{\beta
,\beta^{\prime}}^{(\alpha,\gamma)}G_{\alpha,\beta^{\prime}}(y)F_{\beta
^{\prime},\gamma}(x),\,\,x>y \label{ex}%
\end{equation}
where the R-matrices are determined from admissible braid group
representations. For more on the timelike braid group structure in higher
dimensional conformal QFT we refer to \cite{anomalous}.

Since the AdS-CQFT isomorphism implies a radical reprocessing on the physical
side, it would be interesting to perform the timelike commutation relation
analysis directly within the AdS setting. This has not been done yet.

\section{Generalized Holography in Local Quantum Physics}

The message we can learn from the AdS-conformal correspondence is two-fold. On
the one hand there is the recognition that there are situations where it is
necessary to avoid the use of ``field coordinates'' in favor of directly
working with local algebras. In most concrete situations there were always
convenient field coordinatizations available in terms of which the
calculations simplified. For the AdS-conformal correspondence is however a new
type of problem for which the best way is to stay intrinsic, i.e. to use the
net of algebras.

The second message is that there may exist a holographic relation between
QFT's and their lower dimensional boundaries. We have argued that the degrees
of freedom of $AdS_{d+1}$ are the same as in the corresponding $CQFT_{d}$ on
the boundary even though the Hamiltonians and the associated thermal aspects
are different\footnote{Contrary to a widespread belief, the number of degrees
of freedom of \textit{causally propagating pointlike} AdS$_{d+1}$ theories is
always larger than that of a \textit{causally propagating} conformal theories
CQFT$_{d}$ so that the isomorphism cannot be one among causally propagating
pointlike theories i.e. if the AdS theory is pointlike and causally
propagating, the associated conformal theory has no causal propagation and
hence has to be discarded as unphysical.}. This is the only known case of a
bijection of nets of algebras associated with spacetimes of different
dimensions but with the same maximal spacetime diffeomorphisms group.

Another more frequent kind of holography\footnote{Since our approach tries to
relate the holographic aspects via modular localization ideas to the old
principles of particle physics, we do not have to invoke a new ``holographic
principle''.} occurs for spacetimes with a causal horizon. In that case
certain spacetime diffeomorphisms of the original spacetime act in a ``fuzzy''
nongeometric manner, thus accounting for the fact that the diffeomorphism
group of the horizon is smaller. Let us consider a simple example: the
holographic image of a two-dimensional massive theory in the vacuum
representation restricted to the standard wedge i.e. a Rindler-Unruh
situation. We want to restrict the restrict the d=1+1 wedge algebra
$\mathcal{A}(W)$ to its upper half-line horizon $\mathbb{R}_{+}.$ In a massive
theory we expect that both operator algebras are globally identical
\begin{equation}
\mathcal{A}(W)=\mathcal{A}(\mathbb{R}_{+}) \label{w}%
\end{equation}
although their local net structure is quite different. Classically this
corresponds to the fact that characteristic data on either of the two horizons
determine uniquely the function in the wedge\footnote{This is true in any
dimension. The only exception is d=1+1, mass=0 in which case both horizons are
needed to specify the two chiral components of conformal theories.}. It is
very important to control the data on the entire upper horizon $\mathbb{R}%
_{+}$; in contradistinction to a spacelike interval compact intervals on
$\mathbb{R}_{+}$ do not cast two-dimensional causal shadows. The physical
reason is of course that each point in a small neighborhood below that
interval is in the backward influence cone of some points on $\mathbb{R}_{+}$
which are far removed to the right outside that interval. Only if we take
\textit{all of} $\mathbb{R}_{+},$ we will have $W$ as its two-dimensional
causal shadow$.$

In the general approach to QFT the von Neumann algebra of a compact spacetime
region is, according to the causal shadow property of AQFT (which is a local
version of the time-slice property mentioned in the previous section
\cite{HS}), identical to the algebra of its causally completed region. Each
field theory with a causal propagation (in particular Lagrangian field theory)
fulfills this requirement. If one takes a sequence of spacelike intervals
which approximate a lightlike interval, the causal shadow region becomes
gradually smaller and approaches an interval on the light ray in the limit.
The only way to counteract this shrinking is to extend the spacelike interval
gradually to the right in such a way that the larger lower causal shadow part
becomes the full wedge in the limit.

The correctness of this intuitive idea which suggests the correctness of
(\ref{w}) can be checked against other rigorous results. One rigorous result
from Wigner representation theory (which therefore is limited to free field
theories) together with the application of the Weyl- or CAR- (for halfinteger
spin) functor is the statement that the cyclicity spaces for an interval $I$
on $\mathbb{R}_{+}$ agree with the total space \cite{GLRV}
\begin{equation}
\overline{\mathcal{A}(I)\Omega}=\overline{\mathcal{A}(\mathbb{R}_{+})\Omega
}=\overline{\mathcal{A}(W)\Omega}=H \label{RS}%
\end{equation}
i.e. the validity of the Reeh-Schlieder theorem on the light ray subalgebra.
In fact this holds for all positive energy representations including zero
mass, except zero mass in d=1+1 in which case the decomposition in two chiral
factors prevents its validity. Therefore one only needs to proof the spatial
statement $\overline{\mathcal{A}(\mathbb{R}_{+})\Omega}=\overline
{\mathcal{A}(W)\Omega}=H$ in order to derive (\ref{w}). But this spatial
completeness follows from the causal shadow property for spacelike half-lines
$L$ starting at the origin since the space $\overline{\mathcal{A}(L)\Omega}=H$
and this completeness property cannot get lost in the light ray limit
$L\rightarrow\mathbb{R}_{+}.$ The step from spaces to (\ref{w}) is done with
the help of Takesaki's theorem (mentioned later).

We still have to rigorously define the holographic algebra $\mathcal{A}%
(\mathbb{R}_{+})$ (which turns out to be chiral conformal) and its net
structure $\mathcal{A}(I)$ from $\mathcal{A}(W).$ This is done by the modular
inclusion technique which is one of AQFT most recent mathematical achievements
\cite{Inter}\cite{GLRV}.

The modular way of associating a chiral conformal theory with e.g. a d=1+1
massive theory is the following . Start from the right wedge algebra
$\mathcal{A}(W)$ with apex at the origin and let an upper lightlike
translation $a_{+}$ (which fulfills the energy positivity!) act on
$\mathcal{A}(W)$ and produce an inclusion (all the algebras are von Neumann
algebras)
\begin{equation}
\mathcal{A}(W_{a_{+}})\subset\mathcal{A}(W)
\end{equation}
This inclusion is halfsided ``modular'', i.e. the modular group\footnote{For
presentations of the Tomita-Takesaki modular theory which are close to the
present concepts and notations see \cite{Sch2}. A more extensive presentation
which pays due attention to the importance of modular theory for the new
conceptual setting of QFT is that by Borchers \cite{Borchers}.} $\Delta^{it}$
of ($\mathcal{A}(W),\Omega)$ which is the Lorentz boost acts on $\mathcal{A}%
(W_{a_{+}})$ for $t<0$ as a ``compression''
\begin{equation}
Ad\Delta^{it}\mathcal{A}(W_{a_{+}})\subset\mathcal{A}(W_{a_{+}}),\,\,t<0
\end{equation}
The assumed nontriviality of the net i.e. the intersections\footnote{The
nontriviality of the intersections is in some sense the algebraic counterpart
of the renormalizable short distance behaviour in a quantization approach
which is believed to be required by the mathematical existence of the
Lagrangian theory.} of wedge algebras entails that the relative commutant
(primes on algebras denote their commutant in $B(H))$%
\begin{equation}
\mathcal{A}(W_{a_{+}})^{\prime}\cap\mathcal{A}(W)
\end{equation}
is also nontrivial. Such inclusions are called ``standard''. It is known that
standard modular inclusions correspond to chiral conformal theories, i.e. the
classification problem for the latter is identical to the classification of
all standard modular inclusions \cite{GLW}. In the case at hand the emergence
of the chiral theory is intuitively clear since the only ``living space'' in
agreement with Einstein causality (within the closure of $W$ and spacelike
with respect to the open $W_{a_{+}})$ which one can attribute to the relative
commutant is the lightray interval of length $a_{+}$ starting at the origin.
>From the abstract modular inclusion setting the Hilbert space which the
relative commutant generates from the vacuum could be a subspace $H_{+}\subset
H,$ $H_{+}=PH$ of the original one, but the already mentioned causal shadow
property assures that $H_{+}=H,$ i.e. $P=1$ With the help of the L-boost
(=modular group $\Delta^{it}$ of ($\mathcal{A}(W),\Omega)$ one then defines a
net on the halfline $\mathbb{R}_{+}$ and a global algebra
\begin{align}
\mathcal{A}(\mathbb{R}_{+})  &  =alg\left\{  \cup_{t<o}Ad\Delta^{it}%
(\mathcal{A}(W_{a_{+}})^{\prime}\cap\mathcal{A}(W))\right\}  \subset
\mathcal{A}(W)\\
&  \overline{\mathcal{A}(\mathbb{R}_{+})\Omega}=H,\,\,\curvearrowright
\mathcal{A}(\mathbb{R}_{+})\,=\mathcal{A}(W)\nonumber
\end{align}
The modular group $\Delta^{it}$ of the original algebra leaves this lightray
algebra invariant and hence we are in the situation of the Takesaki theorem
\cite{GLRV}\cite{Borchers} which states that a subalgebra together with the
vacuum which is left invariant by the modular group of the larger algebra, has
modular objects which are restrictions of those of ($\mathcal{A}(W),\Omega)$
and the algebras coalesce iff $\overline{\mathcal{A}(\mathbb{R}_{+})\Omega}%
=H$. The identity (\ref{w}) means that the original modular inclusion was
standard and hence the theory on the light ray is conformal.

The identity of this conformal theory with the massive wedge algebra also
shows that identification of chiral conformal theories with zero mass is a
prejudice. Whether chiral theories are describing massless or massive
situations depends on the identification of the mass operator. In the present
case there exists another second lightlike translation along the lower horizon
and the mass operator is given by the product $P_{+}P_{-}.$ Even though the
spectrum of each $P_{\pm}$ is gapless, as required by conformal invariance,
there is a mass gap in the physical mass operator. Since the lightray algebra
is identical to the wedge algebra, the lower $a_{-}\,$lightray translation
also acts on it; but not as a diffeomorphism but rather in a fuzzy \cite{SW}
i.e. totally nonlocal way relative to the local Moebius group action coming
from the geometry of the upper lightray. So in the lightray representation of
the wedge algebra only the Lorentz boost (which becomes a scale transformation
on $\mathbb{R}_{+}$) and the upper lightray translation are shared as local
diffeomorphism operations in both representations. The \textit{lower lightray
translation is nonlocal} on $\mathcal{A}(\mathbb{R}_{+})$ and the
\textit{Moebius rotation} (after compactification of $\mathbb{R}_{+})$
\textit{is newly created} and acts only partially geometrically on $W.$

It is one of the characteristic features of this generalized holography that
in addition to the local and nonlocal encoding of diffeomorphisms of the
original theory into its lower dimensional holographic image, there are also
partially geometric symmetries as the Moebius rotations transferred back from
the image into the original theory. The degrees of freedom of the chiral
conformal $\mathcal{A}(\mathbb{R}_{+})$ are in some intuitive sense ``more''
than in a standard chiral conformal theory associated with a chiral energy
momentum tensor, because such a standard model would algebraically be to small
in order to carry an additional fuzzy $a_{-}$ lightray translation.

Still another related idea about relations of QFTs on different but this time
equal dimensional spacetimes which uses modular techniques in an essential way
has recently appeared under the name ``Transplantation of Local Nets''
\cite{BMS}.

The present modular inclusion approach to ``lightray physics'', including the
localization and degrees of freedom aspect is another illustration of the
conceptual power of the field coordinate free approach and the modular
inclusion method. In the standard setting there are several fake as well as
genuine (requiring a change of field coordinates) problems with light cone-
restrictions and quantizations. Standard approaches are usually entirely
formal; they tend to overlook localization problems whose understanding is
vital for the physical interpretation of the formalism and furthermore often
use field coordinatizations which become singular on the lightray.

These problems continue in higher dimensions where the wedge horizons are
lightfronts. A typical case which requires new concepts is d=1+2. In that case
the modular method, applied to one wedge, only transfers a small fraction of
the geometric structure of the original theory into a chiral conformal theory
obtained by modular inclusion, which localization-wise should really be
associated with the upper light front horizon of the wedge. The lightfront
quantization (or ``infinite momentum frame'' method) with respect to one
lightfront only cannot account for the full locality informations. Since its
transversal localization remains completely unresolved, the so obtained theory
only contains the longitudinal localization data of a chiral conformal net.

Let me explain the way to get a transversal resolution. In that case one tilts
the wedge by a L-boost which leaves the upper defining light ray for the
initial wedge invariant \cite{Sch1}\cite{Sch2}. One then convinces oneself
that this newly positioned second wedge has a modular associated chiral
conformal theory which, though being unitarily equivalent to the first one,
carries the missing information (which is needed for the reconstruction of the
original d=1+2 theory) in form of its relative position in the common Hilbert
space $H$. The tilted wedge together with the original one can be used to give
a net structure to the original wedge in the transversal direction. Again the
holographic projection of the original net into the horizon has besides
geometric actions also fuzzy and partially local actions.

But instead using the transversal resolution of the 2-dim. horizon for a
constructive approach based on the modular inclusion and intersection method,
it would be somewhat more natural to describe the original theory in terms of
the \textit{two chiral theories} which the modular inclusion associates with
the original wedge and the tilted wedge. In the general d-dimensional case one
would encode the original theory in terms of d-1 copies of one and the same
chiral theory in different positions within one Hilbert space. The name
``chiral scanning'' would hence be more appropriate than holography for such a procedure.

Adding nice names to structural relations is of course by itself not very
constructive. The hope is that by more profound future studies one may develop
criteria which allow a more universal intrinsic algebraic characterization of
those relative positions and chiral theories which allow to construct a
d-dimensional QFT. Chiral theories are the simplest and best understood QFTs
and the study of d-1 copies of them seems to be simpler than to confront
higher dimensional field theories directly.

In fact 't Hooft's original holography \cite{Ho} proposal and Susskind's
\cite{Suss} subsequent work appear much more related to the light front
encoding and/or the related scanning than to the AdS holography with its high
geometric symmetry restriction. The present use of modular inclusions may be
seen as an attempt to find a firmer conceptual and mathematical basis for
those ideas.

The importance of causal horizons in the above considerations suggests to look
for a ``localization entropy'' of causally localized matter as a first step
towards a quantum explanation of the universal Bekenstein area law in black
hole physics. But there is a hurdle right at the start: unlike QM where a
quantization box defines a inside/outside division of the Hilbert space and
the quantum mechanical algebra (type I$_{\infty}$ von Neumann factor) through
a tensor product factorization, the nature of the double cone algebras (the
relativistic causally closed analogs of boxes) in QFT is totally different,
since as hyperfinite type III$_{1}$ von Neumann factors they contain neither
minimal projectors nor are there any pure states among its normal states
\cite{Borchers}. This unusual state of affairs requires the introduction of
the ``split property'' in order to construct the relativistic analogue of the
QM box \cite{Haag}. The physical mechanism behind this property is the strong
vacuum fluctuations of partial charges at the surface of its localization
volume V, one of the oldest and most characteristic phenomena which set apart
QFT from QM.

Let us first try to understand this phenomenon in a mathematically refined
formulation of its original discovery by Heisenberg. Using a smooth spacetime
smearing function consisting of a spatial part $g_{R,\delta}(\mathbf{x})$ with
thickness $\delta$ and localization radius $R$ $\ $multiplied by a compact
support time-smearing $f$ in the definition of the partial charge%

\begin{align}
Q_{R}  &  =\int j_{0}(x)f(x_{0})g_{R,\delta}(\mathbf{x})d^{s}x\\
g_{R,\delta}(\mathbf{x})  &  =\left\{
\begin{array}
[c]{c}%
1,\,\,\left|  \mathbf{x}\right|  <R\\
0,\,\,\left|  \mathbf{x}\right|  >R+\delta
\end{array}
\right. \nonumber\\
f(x_{0})  &  \geq0,\,\,\int f(x_{0})d\,x_{0}=1\nonumber
\end{align}
one finds that the square norm of the partial charge applied to the vacuum
$\left\langle Q_{R}Q_{R}\right\rangle $ diverges with $\delta\rightarrow0$ and
increases for fixed $\delta$ in the limit $R\rightarrow\infty$ as $R^{d-2}$
where d is the spacetime dimension \cite{BDLR}.

But what, if any, could be the message of this area law with that of the would
be localization entropy? We first have to understand the algebraic analogue of
the surface vacuum fluctuation of the partial charges. This turns out to be
the split property i.e. the necessity to work with fuzzy space time boxes in
the form of double cones with a ``collar'' region of thickness $\delta$
separating the inside of the smaller box of radius R from the outside of the
bigger with radius R+$\delta$. In this split situation we do recover the
quantum mechanical inside/outside tensor factorization which refers to a fuzzy
box algebra $\mathcal{N}$ which extends beyond the smaller box into the collar
without sharp geometric boundaries \cite{Haag}. This sets the stage for
defining von Neumann entropy which needs the type I tensor-factorization of
boxes in QM.

There remains however another important difference to Schr\"{o}dinger quantum
mechanics in that the vacuum state remains entangled i.e. does not split into
an inside/outside part but rather remains a highly correlated state with the
Hawking-Unruh temperature. This has paradigmatic consequences for the
conceptual framework of the measurement process in local quantum physics
\cite{Cl-Hal}. It is also the origin of the localization entropy which we have
been looking for. One can show that the vacuum state restricted to the fuzzy
QFT box leads to a nontrivial entropy which diverges with $\delta\rightarrow0$
and increases with the size R of the box in agreement with the above analogy
which intuitively pictures the box entropy of the vacuum as being related to a
partial ``Hamiltonian charge'' via a Gibbs formula in the above sense. As in
that case one also would expect the validity of an entropical area law at
least for large ratios of the diameter divides by the collar size and that the
matter dependence would show up, if not in the coefficient of the area law
itself, at least in its correction terms. The ``Hamiltonian charge'' which we
intuitively relate with a Gibbs formula is not expected to be associated to a
geometrical symmetry but rather to one of the infinitely many
modular-generated fuzzy/hidden symmetries which any QFT possesses. In
particular we find the use of the conformal rotational Hamiltonian which
appeared in the recent literature \cite{Ver} physically ad hoc, especially in
view of the fact that Bekenstein's area law does not require conformal
invariance. Even if in very special conformal situations its spectrum happens
to be similar to that of the logarithm of the modular operator of the
splitting box algebra with respect to the vacuum and the resulting entropy
complies with the Bekenstein area law, such an enigmatic observation will be
helpful only if it leads to a general physical concept; by itself it cannot be
a substitute for a deeper conceptual understanding. The Minkowski analog of
black hole thermodynamics/statistical mechanics in our view is more the
understanding of thermal aspects resulting from (modular) localization rather
than the application of the heat-bath Gibbs formalism.\ 

Various intuitively equivalent forms of localization entropy related to the
split inclusion situation were introduced via the concept of relative entropy
for a pair of states in the work of H. Narnhofer \cite{Narn}. The most
managable version for the purpose of extracting a possible area law which
refers directly to the states seems to be the relative entropy of the vacuum
relative to the ``split vacuum'' on the restricted tensor product algebra
$\mathcal{A}\otimes\mathcal{B}^{\prime},\,\mathcal{A}\subset\mathcal{N}%
,\,\,\mathcal{B}^{\prime}\subset\mathcal{N}^{\prime}\;$where $\mathcal{A}$ is
the smaller double cone algebra and $\mathcal{B}^{\prime}$ the commutant of
the bigger one. There exists \cite{Kosaki} a nice variational formula in terms
of states only for such relative entropy of a von Neumann algebra
$\mathcal{M}$ between two states $\omega_{i},i=1,2$%
\begin{align}
&  S(\omega_{1}|\omega_{2})_{M}=-\left\langle log\Delta_{\omega_{1},\omega
_{2}}\right\rangle _{\omega_{2}}\\
&  =sup\int_{0}^{1}\left[  \frac{\omega(1)}{1+t}-\omega_{1}(y^{\ast
}(t)y(t))-\frac{1}{t}\omega_{2}(x^{\ast}(t)x(t))\frac{dt}{t}\right]
\nonumber\\
&  x(t)=1-y(t),\,\,x(t)\in\mathcal{M}\nonumber
\end{align}

Here $\Delta_{\omega_{1},\omega_{2}}$ is the relative modular operator and for
the case at hand we have to identify $\omega_{1}=\Omega,\omega_{2}%
=\Omega\otimes\Omega$ (the split vacuum) and $\mathcal{M}=\mathcal{A}%
\otimes\mathcal{B}^{\prime}.$ Using some previous nuclearity estimates of
Buchholz and Wichmann \cite{Haag}, Narnhofer carried out a rough estimate for
this entropy and found that it increases \textit{less than the volume} of the
relativistic box of size $R$ \cite{Narn}. In the present setting her result
may be interpreted as a first indication in favor of a Bekenstein area law for
localized quantum matter. I order to obtain more structural inside into this
fundamental and universal phenomenon I started to investigate this problem in
the mathematically more controllable situation of two double cones separated
by a collar in conformal theories \cite{Sch2}. By conformal invariance the
large R behaviour becomes coupled to the short distance behavior in the limit
of vanishing collar size $\delta\rightarrow0.$ One expects to have an easier
conceptual grasp on this ultraviolet behavior as a consequence of the fact
that it reflects truely intrinsic properties of the local algebras and has
nothing to do with short distance divergencies of particular field
coordinates. Besides, conformal theories from an analytic viewpoint are the
simplest theories after free fields. There are as yet no sufficiently concrete
results worthwhile to be reported here.

To avoid misunderstanding, I am not saying that the area law for black holes
is a simple consequence of the area law for localized quantum matter. It would
be a pity if it would, because then not much would be revealed by black hole
physics about the still extremely speculative issue of quantum gravity. Rather
I believe that it is the seemingly very nontrivial conversion of localization
entropy of local quantum physics into the more geometric Killing horizon
entropy of black holes\footnote{The role of the double cone restricted vacuum
in the black hole situation is played by the Hartle-Hawking state restricted
to the outside of the black hole \cite{Wald}.} which will be the crucial step.

\section{Epilogue}

The present analysis of the AdS-CQFT correspondence has its roots in the LSZ
setting of particle physics from which the conformally invariant QFT should
result in the zero mass limit \cite{particles}. The step from the traditional
use of pointlike (Lagrangian) fields to operator algebras indexed by spacetime
regions has been taken a long time ago with the intention to obtain a more
profound understanding of the observed insensitivity of the S-matrix obtained
as the asymptotic limit in the setting of the Lehmann Symanzik-Zimmermann
formalism to changes of field coordinates (``interpolating fields'') within
the (Borchers) equivalence class. This led to a more intrinsic formulation of
QFT called algebraic QFT (AQFT) which relegates the role of fields to a
coordinatization of local algebras in terms of selection of particular
generators. If one wants to use field-coordinatizations alltogether, as was
needed in Rehren's proof of the AdS-CQFT isomorphism, it is appropriate to
avoid the word ``field'' and talk about Local Quantum Physics \cite{Haag}. As
the step from differential geometry with coordinates to the modern intrinsic
coordinate-free formulation did not represent a change of the geometrical
content, one does not change physical principles (but only some concepts for
their implementation) by passing from QFT to LQP. Since certain problems, as
e.g. the abominable short-distance problem in the pointlike formulation
($\sim$coordinate singularities? of which field coordinate?)\footnote{There
are also intrinsic ultraviolet aspects of the local algebras. For example if
one uses the ``split property'' for the definition of the vauum entropy of a
local algebra with a ``collar'' for controlling the vacuum fluctuations near
the causal horizon of the localization region, the entropy diverges with
shrinking collar size in a way which is characteristic for the model but not
for one of its field coordinates \cite{Sch2}.} which always seemed to threaten
the existence of Lagrangian QFT through its long jorney through
renormalization theory, become deemphasized in favor of apparantly different
aspects (ultraviolet divergencies$\rightarrow$nontriviality of certain
intersection of algebras) in the new formulation, this reprocessing of
concepts represents a very healthy change.

The conjecture about the AdS-CQFT \ correspondence comes from string theory.
Although string theory has been the dominant way of thinking in particle
physics publications for at least two decades, its main achievements seems to
be that (with some training and coaching) it allows theoretical physicists to
make contributions to mathematics. Its historical origin in the dual S-matrix
model of Veneziano was very close to the framework of LSZ scattering theory;
in fact it started as a proposal for a nonperturbative crossing symmetric
S-Matrix which fulfilled a very strong form (not suggested by QFT) of crossing
called duality (saturation of crossing on the level of reggeized one-particle
states). This forced the S-matrix to live in a high-dimensional spacetime of
at least 10 dimensions (by invoking another invented structure: supersymmetry).

The next step in the LSZ logic would have been to ask for the understanding of
this high dimensional QFT (i.e. the unique equivalence class of fields or the
local net of algebras) which has this S-matrix as a bona fide physical
S-matrix i.e. as a large time LSZ limit. Unfortunately this never happened;
instead the off-shell transition was performed at a completely different
purely technical place. It was based on the auxiliary observation that the
particle towers which appeared in the lowest order (or lowest genus of Riemann
surfaces which is the analogue of Schwinger's auxiliary eigentime in QFT) can
be reproduced by the mass spectrum of a string. In the original strong
interaction representation of the model this tower was thought to lead to
resonances (poles in the second Riemann sheet) resulting from higher order
interactions destabilizing the higher-lying particles in the tower. It was
this step (which occurred even before the decree of the use of string theory
as a quantum theory of gravity) which is responsible for the lost (and never
recovered since) relation to causality and localization which are the
cornerstones of QFT. Whereas in earlier times quantum field theorist have
thought (without success) about nonlocal alternatives in the form of an
elementary length or a cutoff, recent developments in algebraic QFT have made
abundantly clear that Einstein causality and its strengthed form Haag duality
is inexorably linked with the mathematics of the Tomita-Takesaki modular
theory. This is an extremely deep theory which is able to convert abstract
domain properties of operators and subspaces obtained by applying algebras of
local quantum physics to distinguished state vectors into concrete spacetime
localization geometry (without the necessity to impose any additional
structure from the mathematics of noncommutative geometry). All structural
insight obtained up to now, the charge superselection structure, TCP, braid
group statistics\footnote{Including the appearance of temporal plektonic
structures in higher dimensional conformal field theories mentioned at the end
of the third section.}, V. Jones- as well as the new modular- inclusion theory
mentioned in this paper, the universal nature of holography and the concept of
localized entropy, all these properties depend on the causality aspects of
QFT. So the reasons for giving them up must be very strong (theoretical or
experimental) and amount to much more than the esthetics of differential
geometric consistency observation. The biggest difference to the a more
scholarly and less marketing Zeitgeist of previous times becomes visible if
one looks at the terminology. Whereas e.g. the quasiclassical Bohr-Sommerfeld
theory was presented in a way that left no doubt about its transitory
character and the step towards quantum mechanics was the
\textit{de-mystification} of the quasiclassical antinomies and loose ends,
string theorist often praise their product as a theory of everything and
invite their fellow physicist to read the big latin letter M as ``mystery'' in
a science whose main aim used to be de-mystification.

Having enjoyed the good fortune of proximity to Harry Lehmann to whose memory
I have dedicated this paper, the present crisis often reminds me of good and
healthy times in particle physics when he made his lasting contributions to
particle physics. It would seem to me that in the present absence of profound
experimental discoveries it would be more reasonable and safer to develop
local quantum physics according to its very strong intrinsic logic and
guidance of its underlying physical principles instead of taking off into the
blue yonder under the maxim ``everything goes''.

But apart from a few exceptions there is a lamentable dominance of ideas which
despite their long age have not contributed anything tangible to particle
physics. This danger eminating from this dominance which seriously threatenes
the chance of our most gifted and original young minds to contribute to the
progress of particle physics (and which may even wipe out the very successful
scholarly traditions in the exact sciences altogether) was certainly realized
by the late Harry Lehmann who reacted to it with his characteristc mocking
irony which his friends and collaborators will not forget easily, and which
besides his scientific achievments probably explaines Wolfgang Pauli's
sympathy and support extended to him.

There are indications that members of the older generation (who have been
keeping silence in the face of the mathematical brilliance and exclusiveness
behind some of the present dominant fashion in particle physics) are slowly
becoming aware of the potential danger \cite{To2}\cite{Pen}.

\textbf{Note added}: Although the majority interest has recently shifted away
from the field theoretic AdS-CQFT problem, we find that it serves as an ideal
ilustration how the powerful concepts of AQFT can solve a problem which
otherwise (despite a very large number of papers) would have remained unsolved.

\textbf{Acknowledgement} I am indebted to Gerhard Mack for valuable
suggestions and encouragements.


\begin{thebibliography}{99}
\bibitem{Ma}J. Maldacena, Adv. Theor. Math. Phys. \textbf{2}, (1998) 231

\bibitem{GKP}S. S. Gubser, I. R. Klebanov, A. M. Polyakov, Phys.Lett. B428
(1998) 105

\bibitem{Wi}E. Witten, Adv. Theor. Math. Phys. \textbf{2}, (1998) 253

\bibitem{Frons}C. Fronsdal, Phys. Rev. \textbf{D10}, (1974) 589

\bibitem{AIS}S. J. Avis, C. J. Isham and D. Storey, Phys. Rev. \textbf{D18},
(1978) 3565

\bibitem{Re}K.-H. Rehren, Ann, Henri Poincar\'{e} \textbf{1}, (2000) 607

\bibitem{Ber}M. Bertola, J. Bros, U. Moschella and R. Schaeffer,
\textit{AdS/CFT correspondence for n-point functions, hep-th/9908140}

\bibitem{BFS}D. Buchholz, M. Florig and S. J. Summers, \textit{Hawking-Unruh
temperature and Einstein Causality in anti-de Sitter space-time, }hep-th/9905178

\bibitem{Sch1}B. Schroer, \ Ann. Phys. (N.Y.) \textbf{275}, (1999) 190 and
references therein

B. Schroer, ``Local Quantum Theory beyond Quantization'', \ in \textit{Quantum
Theory and Symmetries}, ed. H.-D. Doebner, V.K Dobrev, J.-D. Henning abd W.
Luecke, World Scientific (2000), hep-th/9912008

\bibitem{Sch2}B. Schroer, J. Math. Phys. \textbf{41}, (2000) 3801

\bibitem{Se}I. E. Segal, \textit{Causality and Symmetry in Cosmology and the
Conformal Group}, Montreal 1976, Proceedings, Group Theoretical Methods In
Physics, New York 1977, 433 and references therein to ealier work of the same author.

\bibitem{LM}M. Luescher and G. Mack, Commun. Math. Phys. \textbf{41}, (1975) 203

\bibitem{HSS}M. Hortacsu, B. Schroer and R. Seiler, Phys. Rev. \textbf{D5},
(1972) 2519

\bibitem{SS}B. Schroer and J. A. Swieca, Phys. Rev. \textbf{D10}, (1974) 480,
\thinspace\ B. Schroer, J. A. Swieca and A. H. Voelkel, Phys. Rev.
\textbf{D11}, (1975) 11

\bibitem{BPZ}A. A. Belavin, A. M. Polyakov and A. B. Zamolodchikov, Nucl.
Phys. \textbf{B247}, (1984), 83

\bibitem{Haag}R. Haag, \textit{Local Quantum Physics}, Springer Verlag (1992)

\bibitem{HS}R. Haag and B. Schroer, J. Math. Phys. \textbf{3}, (1962) 248

\bibitem{anomalous}B. Schroer, \textit{Anomalous Scale Dimensions from
Timelike Braiding}, hep-th/0005134

B. Schroer, \textit{Space- and Time-Like Superselection Rules in Conformal
Quantum Field Theory}, hep-th/0010290

\bibitem{GLRV}D. Guido, R. Longo, J.E. Roberts and R. Verch, \textit{Charged
Sectors, Spin and Statistics in Quantum Field Theory on Curved Spacetimes}, math-ph/9906019

\bibitem{Inter}B. Schroer and H.-W. Wiesbrock, RMP Vol 12 No 2 (Feb 2000) 301-326

\bibitem{SW}B. Schroer and H.-W. Wiesbrock, RMP Vol \textbf{12} No 1 (Jan
2000) 139

B. Schroer and H.-W. Wiesbrock, \textit{Looking beyond the Thermal Horizon:
Hidden Symmetries in Chiral Models}, to appear in RMP Vol 12 No 3 (March 2000)

\bibitem{Suss}L. Susskind, J. Math. Phys. \textbf{36}, (1995) 6377

\bibitem{Ho}G.%
\'{}%
t Hooft, \textit{Dimensional reduction in quantum gravity}, in
Salam-Festschrift, A. Ali et al. eds., World Scientific 1993, page 284

\bibitem{BDLR}D. Buchholz, S. Doplicher, R. Longo and J. H. Roberts, Rev.
Math. Phys. \textbf{Special Issue} (1992) 49

\bibitem{Cl-Hal}Rob Clifton and Hans Halvorson, \textit{Entanglement and open
Systems in Algebraic Quantum Field Theory}, University of Pittsburgh preprint Jan.2000

\bibitem{Ver}E. Verlinde, \textit{On the Holographic Principle in a Radiation
Dominated Universe}, hep-th/0008140.

\bibitem{Narn}H. Narnhofer, in \textit{The State of Matter}, ed. by M.
Aizenman and H. Araki (Wold-Scientific, Singapore) 1994

\bibitem{Kosaki}H. Kosaki, J. Operator Theory \textbf{16}, (1986) 335

\bibitem{Wald}R. M. Wald, \textit{Quantum Field Theory in Cuved Spacetime and
Black Hole Thermodynamics}, University of Chicago Press (1994)

\bibitem{particles}B. Schroer, Phys. Lett. \textbf{B 494,} (2000) 124, \ hep-th/0005110

\bibitem{To2}I. T. Todorov, Two-dimensional conformal field theory and beyond.
Lessons from a continuing fashion, math-phys/0011014

\bibitem{Pen}R. Penrose, \textit{how to compute-help-and hurt scientific
research}, Convergence Winter 1999, page 30

\bibitem{Reh2}K.-H. Rehren, \textit{Local Quantum Observables in the Anti de
Sitter-Conformal QFT Correespondence, }hep-th/0003120

\bibitem{Ruehl}J. Kupsch, W. Ruehl and B. C. Yunn, Ann. Phys. (N.Y.)
\textbf{89}, (1975) 141

\bibitem{GLW}H.-W. Wiesbrock, Lett. Math. Phys. \textbf{31}, (1994) 303,
\ \ \ D. Guido, R. Longo and H.-W. Wiesbrock, Commun. Math. Phys.
\textbf{192}, (1998) \ 217

\bibitem{Borchers}H. J. Borchers, J. Math. Phys. \textbf{41}, (2000) 3604

\bibitem{To1}Nicolay M. Nikolov and Ivan T. Todorov, \textit{\ Rationality of
conformally invariant local correlation functions on compactified Minkowski
space}, hep-th/0009004

\bibitem{BMS}D. Buchholz, J. Mund and S. J. Summers, Transplantation of Local
Nets and Geometric Modular Action on Robertson-Walker Space-Times, hep-th/0011237

D. Buchholz, \textit{Algebraic Quantum Field Theory}, A Status Report hep-th/0011015
\end{thebibliography}
\end{document}